\documentstyle[floats,multicol,aps,epsf,prl]{revtex}

\begin{document}
\draft
\twocolumn[\hsize\textwidth\columnwidth\hsize\csname
@twocolumnfalse\endcsname
\def\btt#1{{\tt$\backslash$#1}}
\title{Plastic Motion of a Flux-line Lattice Driven by Alternating
Current}
\author{W. Henderson$^1$\cite{newadd},   M.J. Higgins$^2$ and E.Y. Andrei$^1$}
\address{$^1$Department of Physics and Astronomy, Rutgers University
Piscataway, New Jersey 08855}
\address{$^2$NEC Research Institute , 4 Independence Way, Princeton, 
         New Jersey 08540}
\maketitle
\begin{abstract}
We have measured the response of the flux-line lattice in the low T$_c$
superconductor, 2H-NbSe$_2$, to finite frequency drives.
In a well-defined range of fields, temperatures, and driving amplitudes
the system exhibits variety of novel non-linear phenomena. Most strikingly,
the flux-lines can move easily in response  to currents that are significantly lower than the DC critical current if the direction of the current is reversed periodically while the amplitude, I, is kept
fixed, but they do not respond to a current that periodically switches
between zero and I, while the direction is kept fixed. Pronounced
memory effects  associated with these phenomena indicate
the presence of dynamically generated structural changes in the flux lattice.
\end{abstract}
\pacs{PACS numbers:}
]

The physics of the flux-line lattice (FLL) 
in low $T_c$ superconductors is governed
by a competition between flux-line interactions, which
favor the formation of an ordered hexagonal lattice,
and  pinning by a random distribution of material defects which favor a
disordered  FLL. 
Recent work on dynamical transitions that occur when the FLL
is driven by an applied current \cite{kosh,fales,giam} has shown that depinning is a complex process that can
involve tearing of the flux lattice and plastic flow
\cite{sb,stan,good,yaron,hnd}. However, while it was shown 
 that the plastic flow is accompanied by an
 increase in signal  noise and that its low frequency response is
anomalous\cite{sb},    little else  is known  about the AC response 
in the plastic flow regime. 

 In this letter, we report the first observation of a rich variety of
effects that occur when the flux lattice is
driven by alternating currents with amplitudes  below the DC
critical current\cite{gord,gord1}, in a region of the phase diagram which was previously associated with plastic flow\cite{sb}. We find that 
the flux-lines move easily when they are shaken back and
forth by a current that periodically reverses direction while
the amplitude, I,  is kept fixed,
but they do not respond to a current that periodically switches
between zero and I, while the direction is kept fixed. We observe  pronounced
memory effects  associated with this effect, which we attribute to dynamically
generated changes in flux lattice structure which are frozen in when the
drive is shut off.   The results appear to be 
 related to plasticity effects which occur in ordinary solids.

\begin{figure}[btp]
\epsfxsize=3.5in
\epsfbox{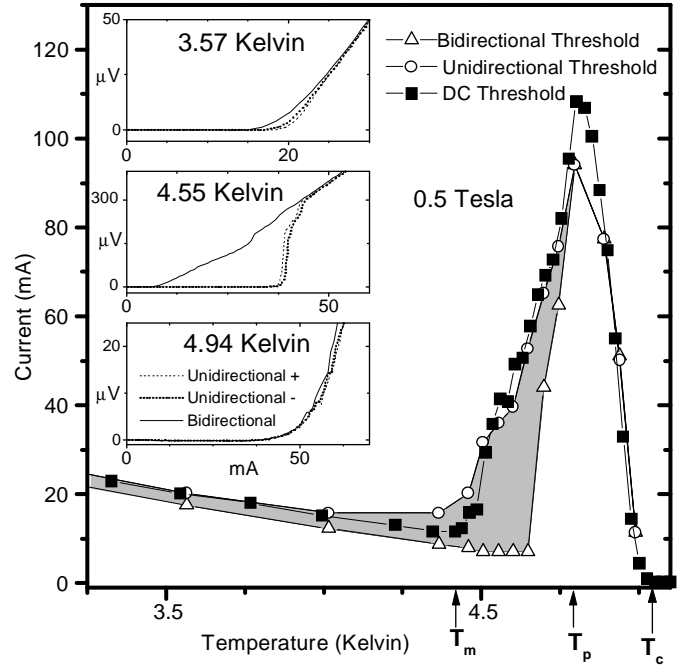}
\protect\caption{ Main Panel: The temperature dependence of the
threshold current at H=0.5 T for unidirectional and bidirectional
pulses and for the DC currents. The lines are guides to the eye.
The insets show the current-voltage characteristic at three
temperatures for each type of pulsed drive.}
\label{fig:fig_pl2}
\end{figure}

  In most weak-pinning superconductors, the
interplay between interactions and disorder leads
to a peak in the critical current, 
$I_c$ which can be observed  as a function of both field
and temperature
\cite{sb,hnd,kes}: 
increasing T at  fixed H leads to  a sharp rise in $I_c$  which sets in at  
   $T=T_m(H)$, reaches its  maximum value at $T_p(H)$, and
finally goes to zero at the transition temperature $T_c(H)$(see Fig. \ref{fig:fig_pl2}).  
Studies of the DC current-voltage relation \cite{sb,hnd},  and
flux-flow noise\cite{marley} 
have revealed that the variation of $I_c$ in the H-T plane is  linked
to the existence of three   states of the FLL  which exhibit  distinct
 dynamic properties. For $T<T_m(H)$
(or equivalently for $H<H_m(T)$), interactions dominate and
the flux-lines form an ordered lattice that responds elastically
 when driven by a current, $I>I_c$. Above $T_p(H)$, the system
is in a glassy state where  the FLL is highly disordered. 
Between $T_m(H)$ and $T_p(H)$, the FLL is in an  intermediate state 
in which it behaves like a soft solid that tears when it is depinned.
The flux motion in this state is thought
 to involve the flow of channels  of relatively weakly pinned flux-lines
past  more strongly pinned neighbors \cite{sb,stan,hnd}.
Such flux-line channels were also 
observed   in numerical simulations \cite{berl,gron,rein,ols}. 
The term plastic flow is  used here to describe  motion in which
different regions of the flux lattice flow with different velocities.

\begin{figure}[btp]
\epsfxsize=3.5in
\epsfbox{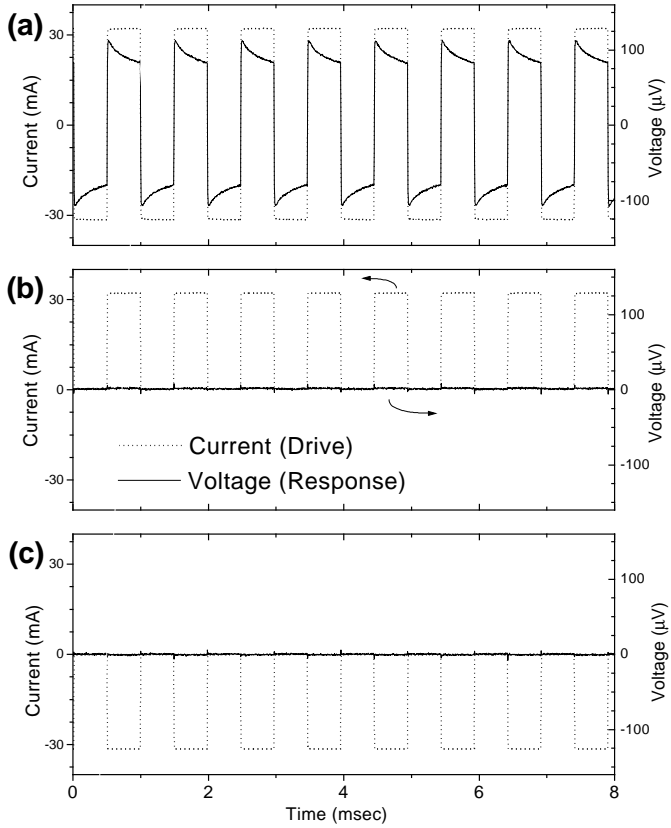}
\protect\caption{The response of the flux lattice at H=.5 T and T=4.59
K to bidirectional pulses (a), positive unidirectional pulses
(b), and negative unidirectional pulses (c).}
\label{fig:fig_pl1}
\end{figure}

The results shown here are for a single crystal
sample (3 x .6 x .015 mm) of the low $T_c$
superconductor
2H-NbSe$_2$. This sample was grown by vapor transport from a
commercial NbSe$_2$ powder containing 200ppm Fe impurities, which
lead to a depressed value of the zero-field transition temperature,
$T_c$
 of 5.8 K. The results reported here were also
observed on two other samples of this material, one of which was grown
from a high purity powder and had a $T_c$ of 7.2K, which is the
nominal value for this material. We focus on results for an impure
sample where  the peak effect is  more pronounced.  
Measurements were done by the standard four lead technique.

Fig. \ref{fig:fig_pl1} shows the response of the flux lattice to three
types  of drives at H=0.5 T and T=4.59 K.
The data were taken with a digital oscilloscope and were
averaged over many cycles to improve the signal to noise ratio. For the
data in Fig. \ref{fig:fig_pl1}a the flux-lines were shaken back and forth
by  a current that switched between  -30mA and
+30mA (bidirectional pulses). The measured voltage, which is
proportional
to the velocity of the flux-lines (averaged over the entire lattice), is
about 30\% of what it would be in the
free flux flow limit (i.e. in the complete absence of pinning).
When the driving current is  switched between 0 and +30mA
or 0 and -30mA (unidirectional pulses),
 the response is
essentially zero (Fig. \ref{fig:fig_pl1}b and \ref{fig:fig_pl1}c).
 The tiny voltage
spikes that coincide with the current switches  are due to pickup
between the
leads and  can be ignored (they are also present when there is no
field and therefore no flux-lines).

A pronounced difference between the response to unidirectional and
bidirectional drives is only observed at certain fields,
temperatures, and driving currents. The insets of
Fig. \ref{fig:fig_pl2} show the current voltage characteristics for
unidirectional and bidirectional pulses at 1KHz for three values of T
at fixed H. The data was obtained 
with a lock-in voltmeter, which measured the root mean square voltage
at the fundamental frequency of the drive. 
The raw data for
the unidirectional drive was multiplied by a factor of 2 to compensate
for the fact that the time during which the current is applied  is half as long as for
the bidirectional case.  
At 4.55 K, the onset of motion for
bidirectional pulses occurs at a threshold current which is a
factor of 5 smaller than for the unidirectional case. For the other two
temperatures shown, there is little or no difference in the response
to the two  kinds of drives.
The region where there is finite response to bidirectional pulses but no response to
unidirectional ones is indicated by the shaded area in  the main panel
of Fig. \ref{fig:fig_pl2}.  Comparison with the DC critical 
currents shows that a large difference between
 the unidirectional and bidirectional thresholds
 only occurs between $T_m(H)$ and $T_p(H)$.

\begin{figure}[btp]
\epsfxsize=3.5in
\epsfbox{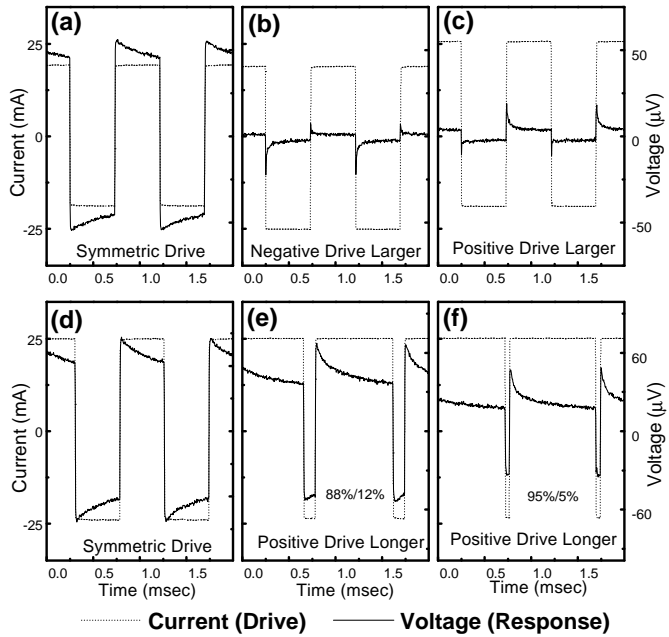}
\protect\caption{The effect of changing the symmetry of the driving
current. In (a) and (d) symmetric bidirectional pulses are applied. In
(b) and (c) the amplitude in one direction is slightly larger. In (e)
and (f), the amplitude is the same in both directions, but the positive
pulses are longer. All data are for T= 4.59 K and H=0.5 T.}
\label{fig:fig_pl3}
\end{figure}

We next studied the effects of varying the
symmetry of the driving current.  The response to changing the 
symmetry of the drive amplitude is shown in Fig. \ref{fig:fig_pl3}a-c: an
asymmetry in the amplitude (i.e. a larger driving current 
in one direction than in the other), 
even when it is small, causes a
significant reduction in the response compared with that of the
symmetric case. Remarkably,
\emph{increasing} the drive in either direction leads to a sharp
\emph{decrease} in the response (provided the drive is not too large).
A similar reduction in the response
is also seen if we add a small DC offset to a symmetric sinusoidal
driving current. But simply increasing the amplitude of a symmetric
bidirectional drive causes the response to increase.

In Fig. \ref{fig:fig_pl3}d-f,  the effects of varying the temporal
symmetry of the  pulse are studied while keeping the pulse amplitude
symmetric. The pulse duration in each direction is varied from
symmetric  (equal pulse lengths)  to  asymmetric (pulse duration  in
one direction up to 20 times longer than in the other) while keeping
the repeat frequency of pulses fixed. In this case the
response is only weakly affected:
there is still a
substantial voltage when the drive is positive (or negative) 95\% of
the time. In this case, there is a   net flow
of flux-lines, not just back and forth motion.
 This behavior is reminiscent of the flow of salt in a
salt shaker, where the occasional reversal of the force is required
to ``unjam'' the system and
maintain the flow. However, this analogy does not work for the
amplitude asymmetry effect.

\begin{figure}[btp]
\epsfxsize=3.5in
\epsfbox{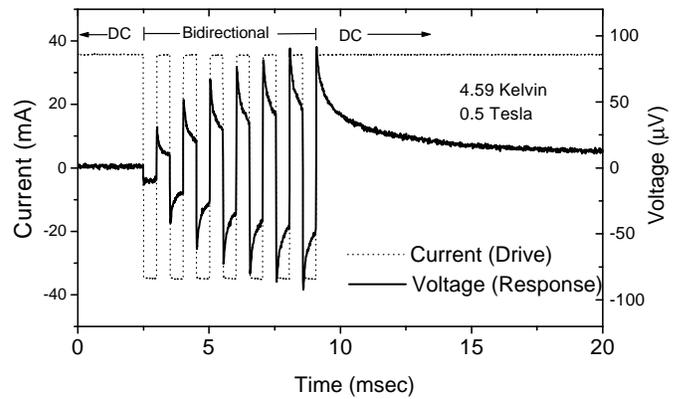}
\protect\caption{In this measurement a DC current was first applied
for several seconds. Then several bidirectional pulses were
applied after which the current was set back to DC. }
\label{fig:fig_pl4}
\end{figure}

The data shown so far were taken after the
response settled to a steady state. If the drive is
switched from the bidirectional pulses shown in Fig. \ref{fig:fig_pl1}a
to the unidirectional pulses in \ref{fig:fig_pl1}b, the response takes
some 
time to decay to zero. We studied such transient
effects  by  observing how the response evolves when the drive is
changed. For the data in Fig. \ref{fig:fig_pl4}, a  +35mA DC current
was first applied for several seconds. Subsequently the current was
switched
back and forth between +35 and -35 mA several times, after which it
was left at +35mA. We see that initially the response is zero,
consistent with the fact that 35mA is less than the DC critical
current. But a small voltage appears as soon as the direction of the
current
is reversed, and this voltage  jumps up to a larger value  on each subsequent
reversal. If the bidirectional drive persists the response  eventually saturates to the steady state shown in Fig. \ref{fig:fig_pl1}a. The
voltage decays somewhat between each reversal.
After the drive is switched back to DC the decay continues
until the response goes back to zero.
The entire pattern shown in the figure repeats exactly, if
the current is cycled repeatedly through the sequence: DC,
bidirectional pulses, DC. 

It is important to note that if the current is set to zero after the
flux-lines are shaken loose with bidirectional pulses, they remain in
an ``easy to move'' state and give a large response as soon as the
current is turned back on. (Waiting times of up to  one hour were
used.) However, when the current is shut off after
applying a DC current, the flux-lines remain in a ``hard to move'' state
and the transient effect shown in Fig. \ref{fig:fig_pl4} will occur if
bidirectional pulses are subsequently applied. The fact that the FLL
``remembers'' how it was driven implies that the bidirectional
 and DC drives cause distinct changes in the structure of the flux lattice.
Interestingly, the manner in which the response decays after the drive is 
switched from AC to DC, depends on the frequency of the drive, indicating that
the frequency plays an important role in determining the FLL structure.

We also studied the  frequency dependence of the  response for symmetric
bidirectional drives (both sinusoidal and square wave), in the range
3Hz-100kHz.
In the region of the phase diagram bracketed by $T_m(H)$ and $T_p(H)$,
for driving
amplitudes between the unidirectional and bidirectional thresholds, the
response is strongly frequency dependent.
The in-phase
response exhibits a crossover between a low frequency regime
in which the response
is vanishingly small and a high frequency regime in which the response
is
a large fraction of the free flux flow value. At the same time  the
out-of-phase response is  negligibly small\cite{lockin}. The
characteristic crossover
frequency is in the range $\sim$1-10kHz and varies with driving
amplitude.
 One could attempt to attribute the  crossover to the motion of
weakly pinned flux-lines elastically coupled to more strongly pinned
ones. However, in this case there would also be a peak (of  comparable
size to the in-phase response at the crossover frequency)
in the frequency dependence of the  out-of-phase response
which would be directly related to the storage and release of elastic
energy during each cycle\cite{kk}. The fact that the  out-of-phase response is
negligibly  small for the experiments discussed here, strongly suggests
that the motion is plastic. 
The phenomena described here  occur  for relatively large currents
( $\geq .2I_c$) only,  where the response is highly nonlinear. By
contrast, when very small driving currents are used,  $I\ll I_c$, 
previous experiments have shown\cite{hnd2}  that the system is linear
and that it exhibits a
crossover in the in-phase response together with the large
peak in the out-of-phase response at  a characteristic ``pinning frequency'',
$\omega _p\sim$ 1-100MHz. This behavior is a direct result of 
the interplay between the flux flow viscosity and \emph{elastic}
coupling between flux-lines and pinning centers.

The results shown above closely
resemble plasticity effects which occur in ordinary solids.
When stresses greater than the ``yield point'' are applied to a solid,
regions of the material slide past one another,
creating permanent plastic deformations.
As plastic flow proceeds in an initially soft material (e.g. annealed copper),
becomes progressively 
more difficult to deform and the motion eventually  stops (provided the applied stress is not too large).
This  phenomenon is known as ``strain hardening'' \cite{hertz}.
The material remains in a hardened state after the applied
stress is removed. However, the yield point after strain hardening is
typically somewhat anisotropic, i.e. less stress is
required to cause plastic flow in the direction opposite to which the
object was deformed during the hardening process. When the stress on 
a heavily strain hardened material is repeatedly reversed, the material
becomes progressively easier to deform: this process is known as
``cyclic softening'' \cite{hertz}. Strain hardening is usually
associated with an increase in the density of
dislocations, whereas cyclic softening involves 
 healing of dislocations and reordering of the lattice.

The behavior of the flux lattice can be interpreted in terms of   these 
plasticity effects. As noted above, there is mounting evidence 
 that  plastic flow in the
FLL  involves the formation of  channels in which the FLL is more
ordered and therefore more weakly pinned than in the surrounding
areas. Thus, our results suggest that symmetric back-and-forth
shaking of flux-lines facillitates  the  formation and growth of these
channels. However, when a unidirectional
drive is applied, the response diminishes quickly, indicating that 
the easy-flow channels become blocked. This decay of the response resembless  the
phenomenon of  strain hardening in ordinary  solids.  
Since the response is small the first time the
direction of the current is reversed after a DC current is applied
(Fig. \ref{fig:fig_pl4}), the strain hardening of the FLL  must be  weakly anisotropic. This is consistent with the observed  strong sensitivity
to the symmetry of the driving amplitudes (Fig. \ref{fig:fig_pl3}).
When the amplitude of the drive is not the same in both directions,
the larger amplitude pulses lead to jamming of the channels which
cannot be completely unjammed  by the smaller amplitude reverse drive.

In summary, we have found that in a well-defined range of fields,
temperatures, and driving amplitudes, the
flux lattice displays  novel types of non-linear response.
In particular, the finite frequency current-voltage characteristics
exhibit threshold currents that are much smaller than the DC
thresholds.  The finite frequency response is extremely sensitive
to asymmetry in the driving amplitude.
However, the response is
 insensitive  to temporal asymmetry in the drive.  The memory
 effects associated
with these results show that dynamically generated changes in the structure
of the flux lattice are involved. As is the case with ordinary solids,
it appears that unidirectional drives tend to disorder the system, while
shaking tends to order it. 
More work is needed,  both theoretical and experimental, in order  to  understand  the mechanism leading to these unusual effects  of jamming, 
softening and plasticity and how they are related to the phase diagram
of  the flux lattice.

Acknowledgments: We thank S. Alexander, S. Cheong, D. Geshkenbein, M. Stephen, 
R. Stinchcomb, R. Walsted,  G. Weng and Z. Xiao for useful discussions;
M. Higgins, S. Bhattacharya, M. Oledzka , J. Sunstrom and M. Greenblatt
for samples.
Work supported by  NSF-DMR-9705389


\begin{references}
\bibitem{lockin}  This follows  directly from   Fig. \ref{fig:fig_pl1}where the response is seen to be in phase with the drive. The in-phase and out-of-phase response were also  
measured with lock-in detection.
\bibitem{kk} In any linear system, the Kramers-Kronig relations
imply that a peak in the frequency dependence of the
out-of-phase response must occur when there is a 
crossover in the  frequency dependence
of the in-phase response.
\end{references}
\end{document}